\documentclass[aps,prl,twocolumn,superscriptaddress,showpacs]{revtex4} 
\newcommand{\figsize}{0.42\textwidth}      
\usepackage{graphicx}
\usepackage{dcolumn}
\usepackage{bm}
\usepackage{setspace}
\def\met{\mbox{${\hbox{$E$\kern-0.6em\lower-.1ex\hbox{/}}}_T$}} 
\def\mex{\mbox{${\hbox{$E$\kern-0.6em\lower-.1ex\hbox{/}}}_x$}} 
\def\mey{\mbox{${\hbox{$E$\kern-0.6em\lower-.1ex\hbox{/}}}_y$}} 
\def\mexy{\mbox{${\hbox{$E$\kern-0.6em\lower-.1ex\hbox{/}}}_{x,y}$}} 
 
\def\D0{D\O}                            
\newcommand{\ptjet}{\rm p_{\rm T}^{\rm jet}}

\newcommand{\ptjetcor}{\rm p_{\rm T, COR}^{\rm jet}}
\newcommand{\kt}{\rm k_{\rm T}}
\newcommand{\yjet}{\rm y^{\rm jet}}
\newcommand{\ptjetcal}{\rm p_{\rm T,CAL}^{\rm jet}}
\newcommand{\ptjetlead}{\rm p_{\rm T,CAL}^{\rm leading \ jet}}
\newcommand{\yjetcal}{\rm y^{\rm jet}_{\rm CAL}}
\newcommand{\fjetcal}{\rm \phi^{\rm jet}_{\rm CAL}}
\newcommand{\ptjethad}{\rm p_{\rm T,HAD}^{\rm jet}}
\newcommand{\yjethad}{\rm y^{\rm jet}_{\rm HAD}}
\newcommand{\fjethad}{\rm \phi^{\rm jet}_{\rm HAD}}
\newcommand{\ETM}       {E_T\hspace{-2.4ex}/\hspace{1.2ex}}

\begin{document}

%
%
\title{
Measurement of the Inclusive Jet Cross Section using the {\boldmath $\rm k_{\rm T}$} algorithm  \\ in 
{\boldmath $p\overline{p}$} Collisions at 
{\boldmath $\sqrt{s}$} = 1.96 TeV\\
}

\affiliation{Institute of Physics, Academia Sinica, Taipei, Taiwan 11529, Republic of China} 
\affiliation{Argonne National Laboratory, Argonne, Illinois 60439} 
\affiliation{Institut de Fisica d'Altes Energies, Universitat Autonoma de Barcelona, E-08193, Bellaterra (Barcelona), Spain} 
\affiliation{Baylor University, Waco, Texas  76798} 
\affiliation{Istituto Nazionale di Fisica Nucleare, University of Bologna, I-40127 Bologna, Italy} 
\affiliation{Brandeis University, Waltham, Massachusetts 02254} 
\affiliation{University of California, Davis, Davis, California  95616} 
\affiliation{University of California, Los Angeles, Los Angeles, California  90024} 
\affiliation{University of California, San Diego, La Jolla, California  92093} 
\affiliation{University of California, Santa Barbara, Santa Barbara, California 93106} 
\affiliation{Instituto de Fisica de Cantabria, CSIC-University of Cantabria, 39005 Santander, Spain} 
\affiliation{Carnegie Mellon University, Pittsburgh, PA  15213} 
\affiliation{Enrico Fermi Institute, University of Chicago, Chicago, Illinois 60637} 
\affiliation{Joint Institute for Nuclear Research, RU-141980 Dubna, Russia} 
\affiliation{Duke University, Durham, North Carolina  27708} 
\affiliation{Fermi National Accelerator Laboratory, Batavia, Illinois 60510} 
\affiliation{University of Florida, Gainesville, Florida  32611} 
\affiliation{Laboratori Nazionali di Frascati, Istituto Nazionale di Fisica Nucleare, I-00044 Frascati, Italy} 
\affiliation{University of Geneva, CH-1211 Geneva 4, Switzerland} 
\affiliation{Glasgow University, Glasgow G12 8QQ, United Kingdom} 
\affiliation{Harvard University, Cambridge, Massachusetts 02138} 
\affiliation{Division of High Energy Physics, Department of Physics, University of Helsinki and Helsinki Institute of Physics, FIN-00014, Helsinki, Finland} 
\affiliation{University of Illinois, Urbana, Illinois 61801} 
\affiliation{The Johns Hopkins University, Baltimore, Maryland 21218} 
\affiliation{Institut f\"{u}r Experimentelle Kernphysik, Universit\"{a}t Karlsruhe, 76128 Karlsruhe, Germany} 
\affiliation{High Energy Accelerator Research Organization (KEK), Tsukuba, Ibaraki 305, Japan} 
\affiliation{Center for High Energy Physics: Kyungpook National University, Taegu 702-701; Seoul National University, Seoul 151-742; and SungKyunKwan University, Suwon 440-746; Korea} 
\affiliation{Ernest Orlando Lawrence Berkeley National Laboratory, Berkeley, California 94720} 
\affiliation{University of Liverpool, Liverpool L69 7ZE, United Kingdom} 
\affiliation{University College London, London WC1E 6BT, United Kingdom} 
\affiliation{Massachusetts Institute of Technology, Cambridge, Massachusetts  02139} 
\affiliation{Institute of Particle Physics: McGill University, Montr\'{e}al, Canada H3A~2T8; and University of Toronto, Toronto, Canada M5S~1A7} 
\affiliation{University of Michigan, Ann Arbor, Michigan 48109} 
\affiliation{Michigan State University, East Lansing, Michigan  48824} 
\affiliation{Institution for Theoretical and Experimental Physics, ITEP, Moscow 117259, Russia} 
\affiliation{University of New Mexico, Albuquerque, New Mexico 87131} 
\affiliation{Northwestern University, Evanston, Illinois  60208} 
\affiliation{The Ohio State University, Columbus, Ohio  43210} 
\affiliation{Okayama University, Okayama 700-8530, Japan} 
\affiliation{Osaka City University, Osaka 588, Japan} 
\affiliation{University of Oxford, Oxford OX1 3RH, United Kingdom} 
\affiliation{University of Padova, Istituto Nazionale di Fisica Nucleare, Sezione di Padova-Trento, I-35131 Padova, Italy} 
\affiliation{LPNHE-Universite de Paris 6/IN2P3-CNRS} 
\affiliation{University of Pennsylvania, Philadelphia, Pennsylvania 19104} 
\affiliation{Istituto Nazionale di Fisica Nucleare Pisa, Universities of Pisa, Siena and Scuola Normale Superiore, I-56127 Pisa, Italy} 
\affiliation{University of Pittsburgh, Pittsburgh, Pennsylvania 15260} 
\affiliation{Purdue University, West Lafayette, Indiana 47907} 
\affiliation{University of Rochester, Rochester, New York 14627} 
\affiliation{The Rockefeller University, New York, New York 10021} 
\affiliation{Istituto Nazionale di Fisica Nucleare, Sezione di Roma 1, University of Rome ``La Sapienza," I-00185 Roma, Italy} 
\affiliation{Rutgers University, Piscataway, New Jersey 08855} 
\affiliation{Texas A\&M University, College Station, Texas 77843} 
\affiliation{Istituto Nazionale di Fisica Nucleare, University of Trieste/\ Udine, Italy} 
\affiliation{University of Tsukuba, Tsukuba, Ibaraki 305, Japan} 
\affiliation{Tufts University, Medford, Massachusetts 02155} 
\affiliation{Waseda University, Tokyo 169, Japan} 
\affiliation{Wayne State University, Detroit, Michigan  48201} 
\affiliation{University of Wisconsin, Madison, Wisconsin 53706} 
\affiliation{Yale University, New Haven, Connecticut 06520} 
\author{A.~Abulencia}
\affiliation{University of Illinois, Urbana, Illinois 61801}
\author{D.~Acosta}
\affiliation{University of Florida, Gainesville, Florida  32611}
\author{J.~Adelman}
\affiliation{Enrico Fermi Institute, University of Chicago, Chicago, Illinois 60637}
\author{T.~Affolder}
\affiliation{University of California, Santa Barbara, Santa Barbara, California 93106}
\author{T.~Akimoto}
\affiliation{University of Tsukuba, Tsukuba, Ibaraki 305, Japan}
\author{M.G.~Albrow}
\affiliation{Fermi National Accelerator Laboratory, Batavia, Illinois 60510}
\author{D.~Ambrose}
\affiliation{Fermi National Accelerator Laboratory, Batavia, Illinois 60510}
\author{S.~Amerio}
\affiliation{University of Padova, Istituto Nazionale di Fisica Nucleare, Sezione di Padova-Trento, I-35131 Padova, Italy}
\author{D.~Amidei}
\affiliation{University of Michigan, Ann Arbor, Michigan 48109}
\author{A.~Anastassov}
\affiliation{Rutgers University, Piscataway, New Jersey 08855}
% ...
\collaboration{CDF Collaboration}
\noaffiliation

%
%
\begin{abstract}

We report on a measurement of the inclusive jet production cross section  in 
$p \overline{p}$ collisions at $\sqrt{s} = 1.96 \ {\rm TeV}$ 
using data collected with
the upgraded Collider Detector at Fermilab in Run II (CDF II) corresponding to an integrated luminosity 
of $385 \ \rm pb^{-1}$.
Jets are reconstructed using the $\kt$ algorithm. 
The  measurement is carried out for jets with rapidity $0.1 < | \yjet | < 0.7$
and transverse momentum in the range \mbox{$ 54 < \ptjet < 700$~GeV/c}. 
The measured cross section is in good agreement with next-to-leading order perturbative QCD predictions 
after the necessary non-perturbative parton-to-hadron  corrections are included.

\end{abstract}

\maketitle



The measurement of  inclusive jet production  
in $p\overline{p}$ collisions at 
$\sqrt{s} = 1.96 \ \rm TeV$  constitutes a  test 
of perturbative QCD (pQCD)~\cite{pqcd} predictions over more than eight orders of magnitude in cross section,  
and is sensitive to the presence of physics beyond the standard model~\cite{newphys}. 
The increased center-of-mass energy and integrated luminosity in Run~II at the Tevatron make it possible to  
measure the cross section for jets with transverse momentum~\cite{coord}, $\ptjet$,  up to about $700 \ {\rm GeV/c}$, extending 
the $\ptjet$ range by more than 150~GeV/c compared with previous results~\cite{runIjet}.
This letter presents a new measurement of the inclusive jet production cross section as a function of $\ptjet$
for jets with $\ptjet > 54 \ \rm GeV/c$ and rapidity~\cite{coord} in the region $0.1 < | \yjet | < 0.7$, where
jets are reconstructed with the  $\kt$ algorithm \cite{ktalgo,soper}.
Similar measurements have been carried out using cone-based jet algorithms in Run II~\cite{runIIjet}. 
However, the $\kt$ algorithm has been widely used for precise QCD measurements at both 
$e^+ e^-$ and $e^\pm p$ colliders, and   
allows a well defined  comparison to the theoretical predictions, 
without introducing  additional  parameters~\cite{soper} in the calculations.
The measurements are corrected 
for detector effects back to the particle (hadron) level~\cite{hadron} and compared to pQCD 
next-to-leading order (NLO)  
predictions~\cite{nlo}. 
A proper comparison with the theory requires corrections to account for non-perturbative  contributions 
that become important at low $\ptjet$. 
This  could  explain the marginal agreement  observed in previous studies~\cite{d0kt}.

The CDF II detector is described in detail elsewhere~\cite{cdfii}.
Here, the sub-detectors most relevant
for this analysis are briefly discussed. The detector has a
charged particle tracking system immersed in a 1.4~T magnetic field, aligned coaxially with the beam line.
A silicon microstrip detector provides tracking over the radial range 1.35 to 28 cm 
and covers the pseudorapidity~\cite{coord} range $|\eta| \leq 2$. 
A cylindrical 3.1 m long open-cell drift chamber covers the radial range from 44 to 132 cm and 
provides full tracking coverage for $|\eta| \leq 1$. Segmented sampling calorimeters, arranged in a projective tower
geometry, surround the tracking system and measure the energy flow of interacting particles in
$|\eta| \leq 3.6$. The central barrel calorimeter~\cite{ccal,wha} covers the region $|\eta| < 1$. 
It consists of electromagnetic and hadronic calorimeters
segmented into  480 towers of size $0.1$ in $\eta$ and  $15^\circ$ in $\phi$.
The measured energy resolution for electrons 
is $\frac{\sigma(E)}{E} = \frac{13.5 \%}{\sqrt{E_T \ [\rm GeV]}} \oplus 2 \% $.
The single-pion energy resolution, as determined from 
test-beam data, is $\frac{50 \%}{\sqrt{E_T \ [\rm GeV]}} \oplus 3 \%$.
A hadronic calorimeter complements the coverage of the central barrel calorimeter in the region
$0.6 < |\eta| < 1.0$ and provides additional forward coverage out to $|\eta| < 1.3$.
The forward region, $1.1 < |\eta| < 3.6$, is covered by scintillator-plate electromagnetic and 
hadronic calorimeters. Cherenkov counters in the region $3.7 < |\eta| < 4.7$ measure the number of inelastic $p \overline{p}$ collisions to 
compute the luminosity~\cite{clc}.


Monte Carlo event samples are used to determine the response of the detector and the
correction factors to the hadron level. The generated samples are passed through a full
CDF detector simulation (based on  {\sc geant3}~\cite{geant} where the {\sc gflash}~\cite{gflash}
package is used to simulate the energy deposition in the calorimeters) and then reconstructed and analyzed using the same 
analysis chain as
for the data. Samples of simulated inclusive jet events have been generated using
the {\sc pythia} 6.203~\cite{pythia} and {\sc herwig} 6.4~\cite{herwig} Monte Carlo generators.
CTEQ5L~\cite{cteq5l} parton distribution functions (PDFs) are used for the proton and antiproton.
The {\sc pythia} samples have been created using a special tuned set of parameters, denoted as
{\sc pythia-tune~a},  that includes  enhanced contributions from initial-state
gluon radiation and  secondary parton interactions between proton/antiproton beam remnants. Tune~A
was developed with dedicated studies of the underlying event 
using the CDF Run I data \cite{underlying}, and  describes the measured jet shapes in Run II~\cite{shapes}. 

The $\kt$ algorithm is used to reconstruct 
jets from the energy depositions in the calorimeter towers with transverse momentum above 0.1 GeV/c. 
First, all towers are considered as protojets. The quantities 
$\rm k_{\rm T,i} = p_{T,i}^2$  and $\rm k_{\rm T, (i,j)} = \rm min(p_{T,i}^2,p_{T,j}^2) \cdot {\Delta R_{i,j}^2}/{D^2}$, 
are computed for each protojet and pair of protojets respectively, where  $\rm p_{T,i}$ denotes the transverse momentum of the 
$\rm i^{\rm th}$ protojet, $\Delta R_{\rm {i,j}}$ is the distance
($y-\phi$ space) between each pair of protojets, and $D$ is a parameter that approximately controls the size of the jet.
All $\rm k_{\rm T,i}$ and $\rm  k_{\rm T,(i,j)}$ values are then collected into a single sorted list. In this combined sorted list, if the smallest quantity is of the type $\rm k_{\rm T,i}$, the corresponding protojet 
is promoted to be a jet
 and removed from the list. Otherwise, if the smallest quantity is of the type  $\rm  k_{\rm T, (i,j)}$, the protojets 
are combined into a single protojet by summing up their four-vector components. The procedure  is iterated over protojets 
until the list is empty.   
The jet transverse momentum, rapidity, and azimuthal angle are denoted as 
$\ptjetcal$, $\yjetcal$, and $\fjetcal$, respectively.
The same jet algorithm is applied to all the final-state particles 
in the Monte Carlo to search for  jets at the hadron level. 
The resulting hadron-level jet 
variables are denoted as $\ptjethad$, $\yjethad$, and $\fjethad$. 


The measurements presented in this letter correspond to a total integrated luminosity of 
$385 \pm 22 \ \rm pb^{-1}$.
Events are selected online using three-level
trigger paths~\cite{trigger} with different prescales. 
In the first-level trigger, a single trigger tower
with $E_T$ above 5 GeV or 10 GeV is required.
In the second-level trigger, clusters are
formed around the selected trigger towers, and 
a cluster
with $E_T$ above 15 to  90 GeV, depending on the trigger path, is required. In the
third-level trigger,  jets are reconstructed using a cone-based algorithm,
   and a jet with $E_T$ above 20 to 100 GeV is required.
Jets are then searched for  using the $\kt$ algorithm, as explained above, with $D=0.7$.
For each trigger data sample, the threshold on the minimum $\ptjetcal$ is chosen in such a way 
that the trigger selection is fully
efficient. 
The events are required to have at least one jet with rapidity in the region $0.1 < |\yjetcal| < 0.7$ and  
corrected transverse momentum (see below) above 54 GeV/c.  The events are selected to have at least 
one reconstructed primary vertex with  $z$-position within 60~cm
around the nominal interaction point. In order to remove beam-related backgrounds and cosmics rays, the events are required to fulfill $\ETM/\sqrt{\Sigma E_T} < F(\ptjetlead)$, where $\ETM$ denotes the missing transverse 
energy~\cite{met}  and $\Sigma E_T = \sum_i E_T^i$ is the total transverse energy of the event, as measured using 
calorimeter towers with $ E_T^i$ above 100~MeV. The threshold function 
$F(\ptjetlead)$ is defined as $F(\ptjet)  = \rm{min} (2+0.0125 \times \ptjet, 7)$, where $\ptjetlead$ is  the uncorrected 
transverse momentum of the leading jet (highest $\ptjet$) and the units are  GeV. This criterion  is designed to 
preserve more than $95 \%$ of the QCD events, as determined from Monte Carlo studies. 
A visual scan for   
$\ptjetcal > 400$~GeV/c showed no remaining backgrounds.

The measured jet transverse momentum includes additional contributions 
from multiple proton-antiproton interactions per bunch crossing at high instantaneous luminosity. 
The data were collected at instantaneous 
luminosities varying between \mbox{$0.2 \times 10^{31} {\rm cm^{-2} s^{-1}}$} and  
$9.6 \times 10^{31} {\rm cm^{-2} s^{-1}}$.
At the highest instantaneous luminosities, an average of 
two interactions per bunch crossing are produced. 
This mainly affects the measured cross section at low 
$\ptjet$ where the contributions are sizeable. 
In CDF, multiple interactions are identified via the presence of additional primary vertices reconstructed from charged particles.
The measured jet transverse momenta are corrected for  this effect by removing a certain amount of transverse momentum, $\epsilon$, for each additional primary vertex. A value $\epsilon  = 1.62^{+ 0.70}_{-0.46}$~GeV/c is determined from the data by requiring that, after the correction is applied,  the 
ratio of cross sections at low and high instantaneous luminosities does not show any $\ptjet$ dependence. 

The reconstruction of the jet variables in the calorimeter  
is studied using Monte Carlo.  These studies
indicate that the angular variables of a jet are
reconstructed with no significant systematic shift and with a resolution better than 0.05 units 
in $y$ and $\phi$ at low $\ptjetcal$, improving as $\ptjetcal$ increases. The measured 
jet transverse momentum systematically underestimates that of the hadron level jet, which is mainly attributed 
to the non-compensating nature of the calorimeter~\cite{calor}.  
For jets with $\ptjetcal$ about 50~GeV/c, the jet transverse
momentum is reconstructed with an average shift of $-19 \%$ and a resolution of $14 \%$. The jet reconstruction 
improves as $\ptjetcal$ increases. For jets with $\ptjetcal$ about 500~GeV/c, the  average shift is  
$-5 \%$ and the resolution is about $7 \%$.  
The bisector method~\cite{bisector} is employed to evaluate
how well the Monte Carlo reproduces the measured jet energy resolutions. Data and 
Monte Carlo agree within a relative uncertainty of $\pm 8 \%$ 
over the whole $\ptjetcal$ range.    

The measured $\ptjetcal$ distribution is unfolded back to the hadron level 
 using Monte Carlo event samples. {\sc pythia-tune~a} provides a
reasonable description of the different jet and underlying event quantities, and 
is used to determine the correction factors in the unfolding procedure. In order to avoid 
any bias on the correction factors due to 
the particular PDF set used,
which translates into  
slightly different simulated $\ptjetcal$ distributions, {\sc pythia-tune~a}  is re-weighted until it accurately follows
the measured $\ptjetcal$ spectrum. The unfolding  is carried out in two steps. 
First, an average correction is computed.
The correlation $< \ptjethad - \ptjetcal >$ vs  $<\ptjetcal>$  
is used to extract correction factors  
which are then applied to the measured jets to obtain the corrected transverse momenta, $\ptjetcor$. 
A raw cross section is defined as
$\frac{{\rm d}^2 \sigma}{{\rm d} \ptjetcor d \yjetcal} = \frac{1}{\mathcal{L}} \frac{N^{\rm jet}_{\rm COR}}{\Delta\ptjetcor \ \Delta\yjetcal }$,
where $N^{\rm jet}_{\rm COR}$ denotes the number of jets in a given $\ptjetcor$ bin,  
$\Delta\ptjetcor$ is the size of the bin, $ \Delta\yjetcal$ 
denotes the region in $\yjetcal$ considered, and  $\mathcal{L}$ is the luminosity.
Second, 
the measurements  are corrected for acceptance
and smearing effects using a bin-by-bin unfolding procedure, which also accounts for
the efficiency of the selection criteria. The unfolding factors,
$U (\ptjetcor) = \frac{d^2 \sigma / d  \ptjethad d \yjethad}{d^2 \sigma / d \ptjetcor d \yjetcal}$,
are extracted from Monte Carlo and      
applied to the measured $\ptjetcor$ distribution to obtain the final result. The factor $U(\ptjetcor)$ increases 
with $\ptjetcor$ and  varies 
between 1.04 at low  $\ptjetcor$  and 1.3 at very high $\ptjetcor$.


A detailed study of the different systematic uncertainties was carried out~\cite{olga}. The measured jet 
energies were varied by  $\pm 2 \%$ at low $\ptjet$ and  $\pm 3 \%$ at high $\ptjet$ 
to account for the uncertainty on the absolute energy scale in 
the calorimeter~\cite{jetcornim}. This introduces an uncertainty in the measured cross section
which varies between $\pm 10 \%$ at low $\ptjet$  and ${}^{+ 55 \%}_{- 40\%}$
 at high $\ptjet$.
A $\pm 8 \%$ uncertainty on the jet energy resolution introduces an uncertainty 
between $\pm 2 \%$ at low $\ptjet$ and $\pm 8 \%$  at high $\ptjet$.    
The unfolding procedure was repeated using {\sc herwig} instead of {\sc pythia-tune~a} to account for the 
uncertainty on the modeling of the parton cascades and the jet fragmentation into hadrons. This translates into an uncertainty about $\pm 5 \%$ at low $\ptjet$.
The unfolding procedure was also carried out using  unweighted 
{\sc pythia-tune~a}, to estimate the residual dependence on the $\ptjet$ spectrum. 
This introduces an uncertainty of $\pm 4 \%$ above 400 GeV/c, which 
becomes negligible at lower $\ptjet$.
%
The quoted uncertainty 
on $\epsilon$ was taken into account.   
The effect on the measured cross section is less than $\pm 3 \%$ and negligible for jets with $\ptjet$ above 200~GeV/c. 
An additional  $5.8 \%$ uncertainty on the total luminosity is  
not included.


%
\begin{figure}[htbp]
\includegraphics[width=\figsize]{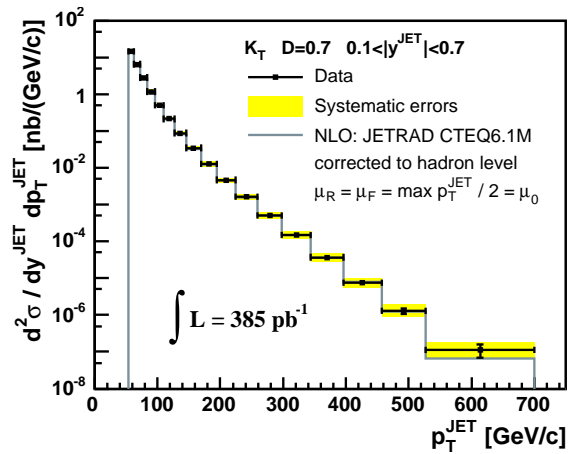}
\caption{Measured inclusive jet cross section (black dots) as a function of $\ptjet$ 
compared to NLO pQCD predictions (histogram). 
The shaded band shows the total systematic uncertainty on the measurement. 
}
\label{fig:pt}
\end{figure}
Figure~\ref{fig:pt} shows the measured cross section as a function 
of $\ptjet$ compared to NLO pQCD predictions. The data are reported in Table~I.
The cross section decreases by more than eight orders of magnitude as $\ptjet$ increases from  54~GeV/c up to 
about 700~GeV/c. 
The NLO pQCD predictions are computed using the {\sc jetrad} program~\cite{nlo} with CTEQ6.1M PDFs~\cite{cteq} and the 
renormalization and factorization scales ($\mu_R$ and $\mu_F$) set to $\mu_0 = \rm max (\ptjet)/2$. 

Different sources of uncertainty in the theoretical 
predictions were considered. 
The main contribution comes from the uncertainty on the PDFs and was computed using  
the Hessian method~\cite{hessian}. It   
varies from ${}^{+20 \%}_{-10 \%}$ at low $\ptjet$, and ${}^{+7 \%}_{-5 \%}$ for $\ptjet$ about 100~GeV/c, to   
${}^{+70 \%}_{-30 \%}$  at  high $\ptjet$, dominated by the limited knowledge of the gluon~PDF. 
An increase of $\mu_R$ and $\mu_F$ from  $\mu_0$ to $2 \mu_0$ reduces the 
theoretical  predictions by $ 2 \%$ at low $\ptjet$ and  
$8 \%$ at high $\ptjet$. Values  significantly smaller than $\mu_0$    
lead to unstable NLO results and were not considered.

The theoretical prediction includes a  
 correction  factor, $C_{\rm HAD}(\ptjet)$,  that approximately accounts for 
non-perturbative contributions
from  the underlying event
and fragmentation into hadrons (see Table~I). 
$C_{\rm HAD}$ was estimated, using
{\sc pythia-tune~a}, as the ratio between the nominal $\ptjethad$ distribution and the one obtained 
by turning off the interactions between proton and antiproton remnants and the fragmentation in the Monte Carlo.   
The parton-to-hadron correction shows a strong $\ptjet$ dependence and decreases as $\ptjet$ increases from
about 1.2 at $\ptjet$ of 54~GeV/c, and 1.1 for $\ptjet$ about 100~GeV/c, to 
1.02 at  high $\ptjet$.
The uncertainty on $C_{\rm HAD}$ is about $13 \%$ at low $\ptjet$ and decreases up to $1.6 \%$ at high $\ptjet$, as 
determined 
using {\sc herwig} instead of {\sc pythia-tune~a}.

\begin{figure}[htbp]
\includegraphics[width=\figsize]{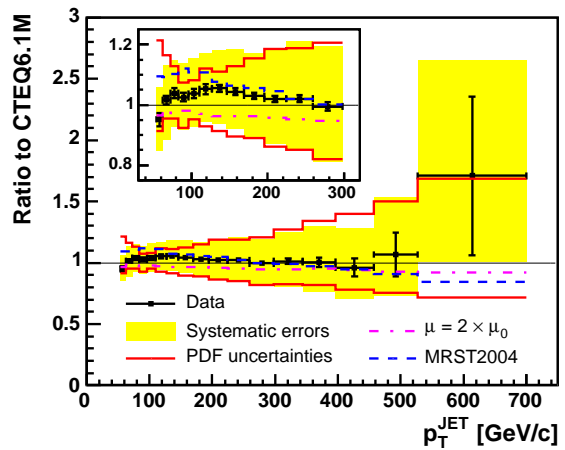}
\caption{Ratio Data/Theory as a function of $\ptjet$. The enclosed figure expands the region $\ptjet < 298$~GeV/c.
The error bars (shaded band)
show the total statistical (systematic) uncertainty on the data.
A $5.8 \%$ uncertainty on the luminosity is
not included.
The solid lines
indicate the PDF uncertainty on the theoretical prediction.
The dashed line presents the ratio of MRST2004  and CTEQ6.1M predictions.
The dotted-dashed line shows the ratio of predictions
with $2 \mu_0$ and $ \mu_0$.
}
\label{fig:ratio}
\end{figure}

Figure~\ref{fig:ratio} shows the ratio data/theory as a function of $\ptjet$. Good agreement is observed 
in the whole range in $\ptjet$. 
A $\chi^2$ test, where all the 
systematic uncertainties on the data are considered independent but fully correlated across different $\ptjet$ bins  
and the uncertainty on $C_{\rm HAD}$ is also included, gives a $\chi^2$ probability of $56 \%$.  
%
In addition, Figure~\ref{fig:ratio} shows the ratio of pQCD predictions 
using MRST2004~\cite{mrst} and CTEQ6.1M PDF sets, 
well inside the 
theoretical and experimental uncertainties.  


\begin{table}[htbp] 
\begin{footnotesize} 
\begin{center} 
\renewcommand{\baselinestretch}{1.2} 
\begin{tabular}{|c|r|c|} \hline
$\ptjet$&  $\frac{\rm{d^2} \sigma}{\rm{d} \ptjet \rm{d} \yjet} \pm {\rm (stat.)} \pm {\rm (sys.)}$ & $C_{\rm HAD} 
\pm {\rm (stat.)} \pm {\rm (sys.)}$ \\  
$[\rm GeV/c]$ & $[\rm nb/(GeV/c)]$ & parton $\rightarrow$ hadron\\ \hline
 54 - 62&   $(      14.6  \pm 0.2   {}^{+ 1.6   }_{- 1.6  }) \times 10^{0\ \ }$ & $ 1.202 \pm 0.013 \pm 0.158$\\   
 62 - 72&   $(      6.53  \pm 0.04  {}^{+ 0.75  }_{- 0.84 }) \times 10^{0\ \ }$ & $ 1.154 \pm 0.003 \pm 0.113$\\ 
 72 - 83&   $(      2.81  \pm 0.02  {}^{+ 0.30  }_{- 0.30 }) \times 10^{0\ \ }$ & $ 1.134 \pm 0.005 \pm 0.094$\\
 83 - 96&   $(      1.18 \pm 0.01 {}^{+ 0.13 }_{- 0.12}) \times 10^{0\ \ }$ & $ 1.113 \pm 0.006 \pm 0.077$\\
 96 - 110&  $(    5.04  \pm 0.04  {}^{+ 0.56  }_{- 0.54  }) \times 10^{-1}$ & $ 1.098 \pm 0.004 \pm 0.066$\\
 110 - 127& $(   2.15  \pm 0.02  {}^{+ 0.25  }_{- 0.22  }) \times 10^{-1}$ & $ 1.079 \pm 0.005 \pm 0.047$\\
 127 - 146& $(   8.81  \pm 0.05  {}^{+ 1.04  }_{- 0.98  }) \times 10^{-2}$ & $ 1.064 \pm 0.003 \pm 0.037$\\
 146 - 169& $(   3.45  \pm 0.02  {}^{+ 0.46  }_{- 0.41  }) \times 10^{-2}$ & $ 1.057 \pm 0.004 \pm 0.030$\\
 169 - 195& $(   1.28 \pm 0.01 {}^{+ 0.17  }_{- 0.17  }) \times 10^{-2}$ & $ 1.047 \pm 0.003 \pm 0.023$\\
 195 - 224& $(   4.67  \pm 0.02  {}^{+ 0.74  }_{- 0.68  }) \times 10^{-3}$ & $ 1.043 \pm 0.003 \pm 0.018$\\
 224 - 259& $(   1.63  \pm 0.01  {}^{+ 0.30  }_{- 0.27  }) \times 10^{-3}$ & $ 1.039 \pm 0.004 \pm 0.015$\\
 259 - 298& $(   5.08  \pm 0.06  {}^{+ 1.02  }_{- 0.93  }) \times 10^{-4}$ & $ 1.034 \pm 0.003 \pm 0.010$\\
 298 - 344& $(   1.50  \pm 0.03  {}^{+ 0.36  }_{- 0.31  }) \times 10^{-4}$ & $ 1.030 \pm 0.005 \pm 0.008$\\
 344 - 396& $(   3.70  \pm 0.14  {}^{+ 1.07  }_{- 0.89  }) \times 10^{-5}$ & $ 1.016 \pm 0.009 \pm 0.006$\\
 396 - 457& $(   7.50  \pm 0.55  {}^{+ 2.52  }_{- 2.01  }) \times 10^{-6}$ & $ 1.017 \pm 0.018 \pm 0.009$\\
 457 - 527& $(   1.31  \pm 0.22  {}^{+ 0.57  }_{- 0.42  }) \times 10^{-6}$ & $ 1.009 \pm 0.003 \pm 0.019$\\
 527 - 700& $(   1.14  \pm 0.43 {}^{+  0.63  }_{- 0.47  }) \times 10^{-7}$ & $ 1.018 \pm 0.002 \pm 0.016$\\ \hline
\end{tabular} 
\end{center}  
\renewcommand{\baselinestretch}{1} 
\end{footnotesize}  
\label{tab:table}   
\caption{Measured inclusive jet differential cross section  as a function of $\ptjet$. 
An additional $5.8 \%$ uncertainty on the luminosity is
not included.
The parton-to-hadron
 correction
factors, $C_{\rm HAD} (\ptjet)$, are applied to the pQCD predictions. 
}
\end{table}

In summary, we have presented results on inclusive jet production 
in $p \overline{p}$ collisions at $\sqrt{s} = 1.96 \ {\rm TeV}$ 
using the $\kt$ algorithm,  for jets with transverse momentum   
$\ptjet > 54$~GeV/c and rapidity in the region $0.1<| \yjet |<0.7$,  
based on $385 \ \rm pb^{-1}$ of CDF Run~II data.
The measured cross section is in agreement with NLO pQCD predictions after 
the necessary non-perturbative parton-to-hadron corrections are taken into account.

We thank the Fermilab staff and the technical staffs of the participating institutions for their vital contributions. This work was supported by the U.S. Department of Energy and National Science Foundation; the Italian Istituto Nazionale di Fisica Nucleare; the Ministry of Education, Culture, Sports, Science and Technology of Japan; the Natural Sciences and Engineering Research Council of Canada; the National Science Council of the Republic of China; the Swiss National Science Foundation; the A.P. Sloan Foundation; the Bundesministerium fuer Bildung und Forschung, Germany; the Korean Science and Engineering Foundation and the Korean Research Foundation; the Particle Physics and Astronomy Research Council and the Royal Society, UK; the Russian Foundation for Basic Research; the Comision Interministerial de Ciencia y Tecnologia, Spain; in part by the European Community's Human Potential Programme under contract HPRN-CT-2002-00292; and the Academy~of~Finland.


\begin{thebibliography}{99}
\bibitem{coord} We use a cylindrical coordinate system about the beam axis in which 
$\theta$ is the polar angle and $\phi$ is the azimuthal angle. We define $E_T = E \ {\rm sin} \theta$,  
$p_T = p \ {\rm sin}  \theta$, $\eta = - {\rm ln(tan(\frac{\theta}{2})) }$, and   
$y = \rm \frac{1}{2} ln(\frac{E+p_z}{E-p_z}) $.
\bibitem{hadron}  The hadron level in the Monte Carlo generators is defined
using all final-state particles with lifetime above $10^{-11} {\rm s}$.
\bibitem{met} $\ETM$ is defined as the norm of $-\sum_i E_T^i \cdot \vec{n}_i$, where $\vec{n}_i$ is the unit vector in the azimuthal 
plane that points from the beamline to the i-th calorimeter tower.
\end{thebibliography}
\end{document}